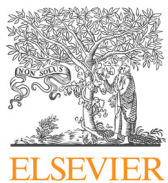
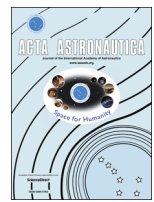

Contents lists available at ScienceDirect

# Acta Astronautica

journal homepage: www.elsevier.com/locate/actaastro

# Robust Mars atmospheric entry integrated navigation based on parameter sensitivity

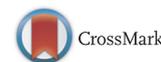

Taishan Lou [a,*], Liangyu Zhao [b]

[a] School of Electric and Information Engineering, Zhengzhou University of Light Industry, Zhengzhou 450002, China
[b] School of Aerospace Engineering, Beijing Institute of Technology, Beijing 100081, China

### ABSTRACT

A robust integrated navigation algorithm based on a special robust desensitized extended Kalman filtering with analytical gain (ADEKF) during the Mars atmospheric entry is proposed. The robust ADEKF is realized by minimizing a new function penalized by a trace weighted norm of the state error sensitivities and giving a closed-form gain matrix. The uncertainties of the Mars atmospheric density and the lift-to-drag ratio are modeled. Sensitivity matrices are defined to character the parameter uncertainties, and corresponding perturbation matrices are introduced to describe the navigation errors with respect to the parameter uncertainties. The numerical simulation results show that the robust integrated navigation algorithm based on the robust ADEKF effectively reduces the negative effects of the two parameter uncertainties and has good consistency during the Mars entry.

© 2015 IAA. Published by Elsevier Ltd. All rights reserved.

## 1. Introduction

On 6th of August 2012, the Curiosity rover of the Mars Science Laboratory successfully landed at a position approximately 2 km from its target, the Gale Crater, with a touchdown ellipse of 19.1 km $\times$ 6.9 km [1]. Although the landing position uncertainties are the smallest during all previous Mars missions, it is not accurate enough for the future Mars exploration missions to have the capability of landing on a high scientific landing site. Factors which contribute to these uncertainties include absence of measurement information, initial entry errors, atmospheric modeling uncertainties, aerodynamic parameter uncertainties and wind drifts during the entry, descent and landing (EDL) phases [2–4]. New high-precision robust navigation EDL technologies, such as high-precision EDL navigation technology, and autonomous hazard detection and avoidance, are necessary for the future Mars pinpoint landing missions [5,6].

The Mars entry phase is the extremely important and dangerous period during the EDL phases, and Mars entry navigation technologies play an important role at a whole precise landing mission [7,8]. However, because optical sensors are blocked by the vehicle's heat shield and plasma sheath, only the inertial measurement unit (IMU) is available during the Mars atmospheric entry. Some researchers presented many technologies, such as adaptive sigma point Kalman filter bank, hierarchical mixture of experts' architecture and multiple model adaptive estimation, to reduce the negative effects of the Mars atmospheric density uncertainties, and try to improve the entry navigation precision only using IMU data [9–11]. However, because of the biases and drifts of IMU, traditional Mars entry vehicles based on IMU dead-reckoning navigation without other external measurement may lead to larger navigation errors [12,13]. Recently, research works showed that the Ultra high frequency radio communication was not blocked by the plasma sheath around the entry vehicle, and provides a new integrated navigation scheme that the external measurements are obtained from the radio communications to improve the online state knowledge during the Mars entry [14,15]. National Aeronautics and Space Administration (NASA) Mars technology program is developing an advanced real-time navigation system using the Electra UHF communication between the Mars entry vehicle and an orbiting satellite or a surface beacon, which is preset on the Mars surface or the previous Mars rover of the last mission [16,17]. Levesque and Lafontaine [4] investigated four innovative measurement scenarios and their observability based on radio ranging during the Mars atmospheric entry, and augmented some parameters into the navigation state vector using the Unscented Kalman filter to improve the navigation precision. The entry initial state errors, the constant bias of IMU, the uncertainties of the Martian atmospheric density are considered during Mars entry navigation using IMU and orbiting/surface radiometric beacons [18–23]. However, for uncertainties of the lift-to-drag ratio (LDR), a very important aerodynamic parameter of the entry vehicle, only a few literatures pay attention on it. Levesque and Lafontaine [4] analyzed the observability of the LDR, and augmented it into the navigation state vector to estimate. EDL simulations of Steinfeldt et al. [24] revealed that altitude of the chute deployment would increase by 1 km as the L/D increased by 35%. New navigation technologies must be developed to adaptively or robustly achieve higher navigation accuracy with the LDR uncertainties.

Desensitized optimal control (DOC) methodology, which is originally presented by Seywald and Kumar [25], has been extended and successfully applied to a wide range of spacecraft optimization problems, such as Mars entry trajectory [26] and Mars pinpoint landing problem of the powered descent phase [27]. The essential idea is to embed a penalty function of the sensitivity with a weighting factor into the original performance index and find an optimal compromise between the sensitivity reduction and the performance. Karlgaard and Shen extended the DOC methodology to the robust filter design problem such that the performance sensitivity of the filters to model uncertain parameters can be reduced [28,29]. The cost function consisting of the posterior covariance trace is penalized by a weighted norm of the state error sensitivities, and desensitized state estimates were obtained by minimizing this cost function. Then, the concept of the desensitized Kalman filter (DKF) was extended to desensitized divided difference filter [30], desensitized unscented Kalman filter [31], in which the cost function was augmented the same penalty function. The effectiveness of DKF has been demonstrated by many applications [20,30]. However, the gain matrix of DKF is obtained by solving a linear equation, not a closed-form solution.

This paper presents a robust integrated navigation algorithm during the Mars atmospheric entry based on a special robust desensitized extended Kalman filter (DEKF) with analytical gain (ADEKF). The robust ADEKF is realized by minimizing a new cost function penalized by a trace weighted norm of the state error sensitivities, which has a scalar sensitivity-weight for each uncertain parameter, and obtaining a closed-form gain matrix. External IMU outputs and radio measurements of the vehicle and orbiting/surface beacons as observations are embedded into the navigation filter. In the above IMU/Radio beacons scheme, the uncertainties of the atmospheric density and the LDR are modeled, and the corresponding sensitivity matrices and perturbation matrices of the vehicle's state estimate errors are employed to describe the effect of the atmospheric density uncertainty and the LDR uncertainty. The robust ADEKF is capable to eliminate the negative effects of initial state errors, atmospheric density uncertainty and LDR uncertainty during the Mars entry, and improves the entry integrated navigation robustness and accuracy.

The paper is organized as follows. Section 2 introduces the Mars entry dynamic model and models the atmospheric density uncertainty and the LDR uncertainty. Section 3 presents the special robust ADEKF based on the sensitivity matrix, and the corresponding perturbation matrix. Section 4 discusses the results of numerical simulations and tests the covariance consistency of the filtering, based on the IMU/Radio integrated navigation. The conclusions are summarized in Section 5.

## 2. Mars entry navigation dynamical system

To overcome the limited available navigation observable measurements during the Mars entry, recent research showed that the ionizing plasma around the entry vehicle can be penetrated by the ultra-high frequency band radio communication [14,17]. A new auto-navigation scheme based on the orbiting/surface radio beacons (also called Mars network) is proposed to improve the observability and entry navigation accuracy for the future Mars landing missions.

Another problem is that the performance of an integrated navigation filter depends largely on the accuracy of the dynamics model. During the Mars entry phase, there are larger uncertainties in the vehicle aerodynamic characteristics, such as the LDR, and the atmospheric density. Unfortunately, it is difficult to precisely model the entry dynamic equations. The parameter uncertainties in the entry dynamic models must be considered. The three degree of freedom dynamics equations are established in the subsequent section.

### 2.1. Dynamics model

For the sake of simplicity, the entry vehicle is assumed to fly in a stationary and quiet atmosphere of a non-rotating planet. The entry dynamics equations of the Mars entry vehicle in the Mars centered Mars-fixed coordinate system are given as [4,32]

$$\dot{r} = v \sin\gamma$$
$$\dot{v} = -D - g(r)\sin\gamma$$
$$\dot{\gamma} = \left(\frac{v}{r} - \frac{g(r)}{v}\right)\cos\gamma + \frac{L}{v}\cos\phi$$
$$\dot{\theta} = \frac{v\cos\gamma \sin\psi}{r\cos\lambda} \qquad (1)$$
$$\dot{\lambda} = \frac{v\cos\gamma \cos\psi}{r}$$
$$\dot{\psi} = \frac{v}{r}\sin\psi\cos\gamma\tan\lambda + \frac{L}{v\cos\gamma}\sin\phi$$

where the altitude $r$ is the distance from the mass center of the entry vehicle to center of the Mars; $v$ is the radial velocity of the entry vehicle; $\gamma$ is the flight path angle (FPA); $\theta$ is the longitude and $\lambda$ is the latitude. The azimuth angle $\psi$ is defined as a clockwise rotation angle starting at due north; and $\phi$ is the bank angle, which is zero in this work. The gravitational acceleration is $g(r) = \mu/r^2$ and $\mu$ is gravitational constant of the Mars. $D$ and $L$ are respectively the aerodynamic drag and lift accelerations given by

$$D = B\bar{q} \qquad (2)$$

$$L = D \cdot L/D \qquad (3)$$

where $L/D$ is the LDR, $\bar{q} = \rho v^2/2$ is the dynamic pressure, and $B = C_D S/m$ is the ballistic coefficient, $C_D$ is the vehicle drag coefficients, $S$ represents the vehicle reference surface area, $m$ denotes the mass of the vehicle, and $\rho$ the Mars atmospheric density.

### 2.2. Measurement model

In this work, an integrated navigation scheme, in which there are one orbiting radio beacon and two predeployed surface radio beacons, is indicated to increase the observation information. The new observation information is the two-way range measurement, which is measured by the radio communication between the vehicle and an orbiting radio satellite or a surface radio beacon (such as the previous landers on the Mars) within sight [33]. The possible radio beacon sources for the Mars entry integrated navigation are shown in Fig. 1. Of course, the IMU will provide three components of acceleration and attitude rates during the Mars entry.

#### 2.2.1. IMU measurement

The accelerometers of the IMU measure the specific force components along three orthogonal axes. Three components of acceleration measured by the accelerometers are provided by

$$\tilde{\mathbf{a}} = \mathbf{a} + \mathbf{b}_a + \boldsymbol{\eta}_a \qquad (4)$$

where $\tilde{\mathbf{a}}$ is the accelerometer output along body axes, $\mathbf{a}$ is the true linear acceleration, $\mathbf{b}_a$ is the acceleration bias, and $\boldsymbol{\eta}_a$ is the zero-mean white Gaussian noise.

In this study, the accelerometer measurement model is defined as

$$\mathbf{a} = \begin{bmatrix} -D & -D \cdot L/D \cdot \sin\phi & D \cdot L/D \cdot \cos\phi \end{bmatrix}^T \qquad (5)$$

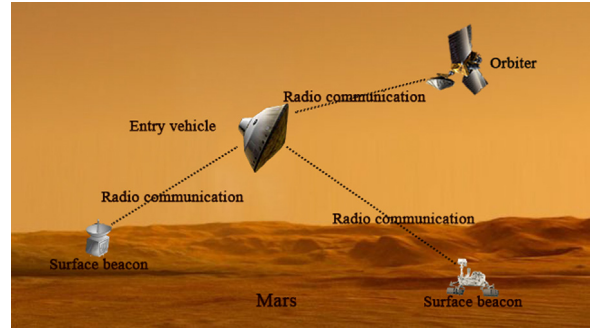

**Fig. 1.** Diagram of IMU/Radio measurement based on integrated navigation.

#### 2.2.2. Two-way range measurement

The two-way range measurement provides the distance between the entry vehicle and an orbiting radio satellite or a surface radio beacon [33]. The two-way range measurement $\tilde{R}$ is reconstructed by

$$\tilde{R}_k = \sqrt{(\mathbf{r}_l - \mathbf{r}_k)^T(\mathbf{r}_l - \mathbf{r}_k)} + \xi_R \qquad (6)$$

where $\mathbf{r}_l$ is the position vector of entry vehicle; $\mathbf{r}_k$ is the position vector of orbiting radio satellite ($k = o$) or surface radio beacon ($k = b$); and $\xi_R$ is the range noise with zero-mean Gaussian white noise.

### 2.3. Martian atmosphere uncertainty model

The Martian atmosphere model is very important for the Mars entry dynamic model, and directly affects the landing precision of the vehicle. The Martian atmosphere model uncertainty mainly comes from the density, wind, dust content etc. The NASA Marshall Space Flight Center gave a high-fidelity Mars-GRAM atmospheric model considering the variation of temperature, pressure, density and wind components with height, latitude, time of day and celestial longitude of the Sun [34]. The Committee on Space Research (COSPAR) established the COSPAR model of the Martian atmosphere, which based on the interpolation of the Viking data, and it is often used as a time invariant approximate atmospheric model in the filters. However, these models only can simply forecast the Martian atmosphere and density of the upper atmosphere, but inaccurately.

For the great uncertainty, the effectiveness of the complex models is degraded, and an approximate exponential Mars atmospheric density model in this work is assumed as follows

$$\rho = \rho_0 \exp\{(r_0 - r)/h_s\} \qquad (7)$$

where $\rho_0$ denotes the nominal reference density on the surface of the Mars, $r_0$ is the nominal reference radial position ($r_0 = 3437.2$ km), and $h_s$ is the constant scale height ($h_s = 7.5$ km).

So the density uncertainty is described as

$$\rho_0 = \bar{\rho}_0 + c_1 \times \bar{\rho}_0 \qquad (8)$$

where $\bar{\rho}_0$ is the nominal reference density, and $c_1$ is the percentage of the atmosphere density uncertainty.

## 2.4. Lift-to-drag ratio uncertainty model

The LDR is very important for the choice of landing sites and the entry atmospheric zone of the vehicle from the orbit. The choice of landing sites is constrained to surface within a hundred kilometers by using a low LDR vehicle, and a high LDR vehicle can enlarge longitudinal and lateral ranges of the landing site and enhanced the mission flexibility [35].

The LDR is defined as the ratio of the lift coefficient $C_L$ to the drag coefficient $C_D$, and its general expression is

$$L/D = \frac{C_L}{C_D} \tag{9}$$

In general, $C_L$ and $C_D$ are not constant for the vehicle flight usually goes through both hypersonic and supersonic flight regimes. Design for simplicity, the aerodynamic coefficients $C_L$ and $C_D$ will be assumed to be constants in filtering implementation, although the LDR will be at its peak value during the dynamic part of the entry.

For the pinpoint landing mission, the LDR uncertainty should be considered, and the uncertainty propagation law should be researched. The percentage uncertainty of the LDR is modeled as

$$L/D = \overline{L/D}_0 + c_2 \cdot \overline{L/D}_0 \tag{10}$$

where $\overline{L/D}_0$ is the nominal reference LDR, and $c_2$ is the percentage of the LDR uncertainty.

## 2.5. Mars entry navigation system

During the Mars entry phase, $\mathbf{x} = \begin{bmatrix} r & \theta & \lambda & v & \gamma & \psi \end{bmatrix}^T$ is defined as the state variables of the entry vehicle. The dynamic model in Eq. (1) is rewritten with the noise term $\mathbf{w}(t)$ as follows:

$$\mathbf{x}(t) = \mathbf{f}(\mathbf{x}(t), \mathbf{c}, t) + \mathbf{w}(t) = \begin{bmatrix} v \sin\gamma \\ -D - g(r)\sin\gamma \\ \left(\frac{v}{r} - \frac{g(r)}{v}\right)\cos\gamma + \frac{L}{v}\cos\phi \\ \frac{v\cos\gamma\sin\psi}{r\cos\lambda} \\ \frac{v\cos\gamma\cos\psi}{r} \\ \frac{v}{r}\sin\psi\cos\gamma\tan\lambda + \frac{L}{v\cos\gamma}\sin\phi \end{bmatrix} + \mathbf{w}(t), \tag{11}$$

and the measurement model with noise $\mathbf{v}(t)$ is constructed as

$$\mathbf{z}(t) = \mathbf{h}(\mathbf{x}(t), \mathbf{c}, t) + \mathbf{v}(t) = \begin{bmatrix} \mathbf{a} & \tilde{R}_1 & \tilde{R}_2 & \tilde{R}_3 \end{bmatrix}^T \tag{12}$$

where $\mathbf{c} = [c_1, c_2]^T$ denotes the uncertain parameter vectors, the atmosphere density and the LDR, in the dynamic model, and $\tilde{R}_i (i=1,2,3)$ the two-way range measurements.

In the subsequent integrated navigation the above dynamical models can be discretized as follows

$$\mathbf{x}_k = \mathbf{f}(\mathbf{x}_{k-1}, \mathbf{c}, t_{k-1}) + \mathbf{w}_{k-1} \tag{13}$$

$$\mathbf{z}_k = \mathbf{h}(\mathbf{x}_k, \mathbf{c}, t_{k-1}) + \mathbf{v}_k \tag{14}$$

where $\mathbf{w}_k$ and $\mathbf{v}_k$ are independent zero-mean Gaussian noise processes, and their covariance are respectively $\mathbf{Q}_k$ and $\mathbf{R}_k$. They satisfy

$$E[\mathbf{w}_k \mathbf{w}_j^T] = \mathbf{Q}_k \delta_{ij}, E[\mathbf{v}_k \mathbf{v}_j^T] = \mathbf{R}_k \delta_{ij}, E[\mathbf{w}_k \mathbf{v}_j^T] = \mathbf{0} \tag{15}$$

where $\delta_{kj}$ is the Kronecker delta function, and $\mathbf{Q}_k$ is positive semi-definite, and $\mathbf{R}_k$ is positive definite.

## 3. Robust desensitized extended Kalman filtering based on parameter sensitivity

Without loss of generality, consider the discrete nonlinear process and measurement models given by

$$\mathbf{x}_k = \mathbf{f}(\mathbf{x}_{k-1}, \mathbf{c}, \mathbf{w}_{k-1}, t_{k-1}) \tag{16}$$

$$\mathbf{z}_k = \mathbf{h}(\mathbf{x}_k, \mathbf{c}, \mathbf{v}_k, t_k) \tag{17}$$

where $\mathbf{x}_k$ is the $n \times 1$ state vector, and $\mathbf{z}_k$ is the $m \times 1$ measurement vector. $\mathbf{f}$ is the dynamic vector-valued function, $\mathbf{h}$ is the nonlinear measurement vector-valued function. $\mathbf{c}$ is referred to as the $\ell \times 1$ uncertain parameter vector. $\mathbf{w}_k$ and $\mathbf{v}_k$ satisfy the conditions in Eq. (15).

In this work, the uncertain model parameter vector is assumed as a pre-estimate $\mathbf{c} = \overline{\mathbf{c}}$, with the a priori knowledge.

### 3.1. Extended Kalman filter

In the conventional extended Kalman filter, the state time update equations are

$$\hat{\mathbf{x}}_k^- = \mathbf{f}\left(\hat{\mathbf{x}}_{k-1}^+, \overline{\mathbf{c}}, 0, t_{k-1}\right) \tag{18}$$

$$\mathbf{P}_k^- = E[\mathbf{e}_k^- \mathbf{e}_k^{-T}] = \mathbf{\Phi}_{k/k-1} \mathbf{P}_{k-1}^+ \mathbf{\Phi}_{k/k-1}^T + \mathbf{\Gamma}_{k-1} \mathbf{Q}_{k-1} \mathbf{\Gamma}_{k-1}^T \tag{19}$$

where the superscripts $^-$ denote a priori and $^+$ denote a posteriori, and the overbar indicates the corresponding estimate function of the parameter. $\mathbf{\Phi}_{k/k-1} = \frac{\partial \mathbf{f}(\mathbf{x}_{k-1}, \overline{\mathbf{c}}, \mathbf{w}_{k-1}, t_{k-1})}{\partial \mathbf{x}_{k-1}}\bigg|_{\mathbf{x}_{k-1} = \hat{\mathbf{x}}_{k-1}^+}$ and $\mathbf{\Gamma}_k = \frac{\partial \mathbf{f}(\mathbf{x}_k, \overline{\mathbf{c}}, \mathbf{w}_k, t_k)}{\partial \mathbf{w}_k}\bigg|_{\mathbf{x}_k = \hat{\mathbf{x}}_k^+}$ are the coefficient matrices. $\mathbf{P}_k^-$ is the error covariance propagation matrix, and $\mathbf{e}_k^- = \hat{\mathbf{x}}_k^- - \mathbf{x}_k$ is priori estimation error.

The measurement update equations are

$$\hat{\mathbf{x}}_k^+ = \hat{\mathbf{x}}_k^- + \mathbf{K}_k[\mathbf{z}_k - \mathbf{h}(\hat{\mathbf{x}}_k^-, \overline{\mathbf{c}}, 0, t_k)] \tag{20}$$

$$\mathbf{P}_k^+ = E[\mathbf{e}_k^+ \mathbf{e}_k^{+T}] = (\mathbf{I} - \mathbf{K}_k \mathbf{H}_k)\mathbf{P}_k(\mathbf{I} - \mathbf{K}_k \mathbf{H}_k)^T + \mathbf{K}_k \mathbf{\Upsilon}_k \mathbf{R}_k \mathbf{\Upsilon}_k^T \mathbf{K}_k^T \tag{21}$$

$$\mathbf{K}_k = \mathbf{P}_k^- \mathbf{H}_k^T (\mathbf{H}_k \mathbf{P}_k^- \mathbf{H}_k^T + \mathbf{\Upsilon}_k \mathbf{R}_k \mathbf{\Upsilon}_k^T)^{-1} \tag{22}$$

where $\mathbf{H}_k = \frac{\partial \mathbf{h}(\mathbf{x}_k, \overline{\mathbf{c}}, \mathbf{v}_k, t_k)}{\partial \mathbf{x}_k}\bigg|_{\mathbf{x}_k = \hat{\mathbf{x}}_k^-}$ and $\mathbf{\Upsilon}_k = \frac{\partial \mathbf{h}(\mathbf{x}_k, \overline{\mathbf{c}}, \mathbf{v}_k, t_k)}{\partial \mathbf{v}_k}\bigg|_{\mathbf{x}_k = \hat{\mathbf{x}}_k^-}$ are the coefficient matrices. $\mathbf{P}_k^+$ is the state estimate error covariance matrix, and $\mathbf{e}_k^+ = \hat{\mathbf{x}}_k^+ - \mathbf{x}_k$ is the posteriori estimation error. Eq. (21) is valid for any $\mathbf{K}_k$. The Kalman optimal gain $\mathbf{K}_k$ is obtained by minimizing the cost function $J = Tr(\mathbf{P}_k^+)$, in which "$Tr$" denotes the trace of the matrix, and the result is Eq. (22).

Under the basic assumptions of the Kalman filter (no model and parameter uncertainty, zero-mean white-noise sequence, known process and measurement models, etc.), the state estimate are unbiased. It means that the estimation errors of the Kalman filter satisfy

$$E[\boldsymbol{e}_k^-] = 0, E[\boldsymbol{e}_k^+] = 0 \quad (23)$$

### 3.2. Sensitivity matrix and perturbation matrix

For the model parameter uncertainties, the basic assumptions of the Kalman filter cannot be satisfied, and the state estimates may be biased and even divergence. So, the state estimate error sensitivities of the parameter vector $\boldsymbol{c}$ could be defined as [31,36]

$$\boldsymbol{S}_k^- = \frac{\partial \boldsymbol{e}_k^-}{\partial \boldsymbol{c}} = \frac{\partial \hat{\boldsymbol{x}}_k^-}{\partial \boldsymbol{c}} \quad (24)$$

$$\boldsymbol{S}_k^+ = \frac{\partial \boldsymbol{e}_k^+}{\partial \boldsymbol{c}} = \frac{\partial \hat{\boldsymbol{x}}_k^+}{\partial \boldsymbol{c}} \quad (25)$$

where $\boldsymbol{S} = (\boldsymbol{s}_1, \boldsymbol{s}_2, \cdots, \boldsymbol{s}_\ell)$ is the $n \times \ell$ sensitivity matrix, $\boldsymbol{s}_i = \partial \boldsymbol{x} / \partial c_i$ denotes the sensitivities of the state estimate $\boldsymbol{x}$ to the $i$th component of the parameter vector $\boldsymbol{c}$. Note that the sensitivity of the true state is $\partial \boldsymbol{x} / \partial \boldsymbol{c} = 0$ in Eqs. (24) and (25) because the true state does not vary with the assumed value of parameter vector $\boldsymbol{c}$.

From the basic Eqs. (18) and (20) of the conventional Kalman filter, the propagation equations can be obtain by

$$\boldsymbol{S}_k^- = \boldsymbol{\Phi}_{k/k-1} \boldsymbol{S}_{k-1}^+ + \boldsymbol{\Psi}_{k-1} \quad (26)$$

$$\boldsymbol{S}_k^+ = \boldsymbol{S}_k^- - \boldsymbol{K}_k \boldsymbol{\gamma}_k \quad (27)$$

where $\boldsymbol{\Psi}_{k-1} = \frac{\partial \boldsymbol{f}(\hat{\boldsymbol{x}}_{k-1}^+, \boldsymbol{c}, \boldsymbol{w}_{k-1}, t_{k-1})}{\partial \boldsymbol{c}} |_{\boldsymbol{c} = \bar{\boldsymbol{c}}}$, and

$$\boldsymbol{\gamma}_k = \boldsymbol{H}_k \boldsymbol{S}_k^- + \boldsymbol{H}_k^c \quad (28)$$

where $\boldsymbol{H}_k^c = \frac{\partial \boldsymbol{h}(\hat{\boldsymbol{x}}_k^-, \boldsymbol{c}, \boldsymbol{v}_k, t_k)}{\partial \boldsymbol{c}} |_{\boldsymbol{c} = \bar{\boldsymbol{c}}}$. Note that the sensitivity of gain matrix is assumed as $\partial \boldsymbol{K} / \partial \boldsymbol{c} = 0$ in Eq. (27), because any $\partial \boldsymbol{K} / \partial \boldsymbol{c} \neq 0$ means that the solution for the optimal gain is a function of the residual, which violates the basis for the linear update equation given in Eq. (20) [28,31].

The sensitivity matrix describes how the state estimate $\hat{\boldsymbol{x}}$ varies with the uncertain parameter vector $\boldsymbol{c}$. To evaluate the effect of the parameter uncertainties on the state estimate error, the perturbation matrix is defined by

$$\boldsymbol{\Gamma} = \boldsymbol{S} \cdot \boldsymbol{\Sigma}_c \quad (29)$$

where $\boldsymbol{\Sigma}_c = \text{diag}[\sigma_{c_1}, \sigma_{c_2}, \cdots, \sigma_{c_\ell}]$ denotes the diagonal matrix, in which the elements are the standard deviation $\sigma_{c_i}(i = 1, 2, \cdots, \ell)$ of each parameter. Each element, $\Gamma_{ij}$, of the perturbation matrix $\boldsymbol{\Gamma}$ gives the error in the state estimate $\hat{\boldsymbol{x}}$ due to $1\sigma$ error in each uncertain parameter $c_j$.

### 3.3. Robust desensitized extended Kalman filtering with analytical gain

To eliminate the negative effects of the parameter uncertainties, Karlgaard and Shen [28,29] introduced the desensitized optimal control methodology into the filtering, and proposed a desensitized optimal filtering to mitigate these negative effects based on a new cost function including the state error sensitivities. They defined the state error sensitivities $\boldsymbol{s}_{i_k}^- = \partial x_k^- / \partial c_i$ and $\boldsymbol{s}_{i_k}^+ = \partial x_k^+ / \partial c_i$ to each parameter $c_i$, and augmented the cost function consisting of the posterior covariance trace with a penalty function consisting of a weight normal of the a posteriori sensitivity by

$$J = Tr(\boldsymbol{P}_k^+) + \sum_{i=1}^{\ell} \boldsymbol{s}_{i,k}^{+T} \boldsymbol{W}_i \boldsymbol{s}_{i,k}^+ \quad (30)$$

where $\boldsymbol{W}_i$ is a $n \times n$ user-specified symmetric positive semi-definite sensitivity–weighting matrix for the $i$th sensitivity. Here, the sensitivity–weighting matrix $\boldsymbol{W}_i$ is a time-variant tuning parameter, which makes a balance between the possibly competing objectives of minimizing the variance and minimizing the sensitivity [30].

Then, taking the derivative with respect to $\boldsymbol{K}_k$, and setting $\partial J / \partial \boldsymbol{K}_k = 0$, yields a linear matrix equation about the gain matrix $\boldsymbol{K}_k$

$$\boldsymbol{K}_k(\boldsymbol{H}_k \boldsymbol{P}_k^- \boldsymbol{H}_k^T + \boldsymbol{\Upsilon}_k \boldsymbol{R}_k \boldsymbol{\Upsilon}_k^T) + \sum_{i=1}^{\ell} \boldsymbol{W}_i \boldsymbol{K}_k \boldsymbol{\gamma}_{i,k} \boldsymbol{\gamma}_{i,k}^T = \boldsymbol{P}_k^- \boldsymbol{H}_k^T + \sum_{i=1}^{\ell} \boldsymbol{W}_i \boldsymbol{s}_{i,k}^- \boldsymbol{\gamma}_{i,k}^T \quad (31)$$

and solving the linear equation algebraically will obtain the gain matrix $\boldsymbol{K}_k$. This implies that the cost of the computational power and the required processing time will increase significantly, especially when the dimension of the state is large.

In this work, based on the sensitivity matrices in Eqs. (26) and (27), which are different from the definitions in DEKF, a special robust desensitized extended Kalman filter with analytical gain is presented. This special DEKF simplifies the sensitivity–weighting matrix of the uncertain parameters by substituting the $n \times n$ matrix $\boldsymbol{W}_i$ as a scalar, and this makes that the gain matrix obtained by solving a linear matrix equation is instead of a closed-form solution.

A new cost function based on the posterior covariance trace and a trace penalty function consisting of the weighted norm of the posterior sensitivity matrix is redefined as

$$J_a = Tr(\boldsymbol{P}_k^+) + Tr(\boldsymbol{S}_k^+ \boldsymbol{W} \boldsymbol{S}_k^{+T}) \quad (32)$$

where $\boldsymbol{W}$ is a $\ell \times \ell$ symmetric positive semi-definite weighting matrix for the uncertain parameters. This sensitivity–weighting matrix is a diagonal matrix, in which each element on the main diagonal is the scalar sensitivity–weighting of each uncertain parameter for all the states of the dynamic model.

Substituting Eqs. (21) and (27) into Eq. (32) and taking the derivative with respect to the gain $\boldsymbol{K}_k$, and using the trace derivative properties in Appendix A, yields

$$\frac{\partial J_a}{\partial \boldsymbol{K}_k} = \frac{\partial}{\partial \boldsymbol{K}_k} Tr(\boldsymbol{P}_k^+) + \frac{\partial}{\partial \boldsymbol{K}_k} Tr(\boldsymbol{S}_k^+ \boldsymbol{W} \boldsymbol{S}_k^{+T})$$
$$= 2\boldsymbol{K}_k \left( \boldsymbol{H}_k \boldsymbol{P}_k^- \boldsymbol{H}_k^T + \boldsymbol{\Upsilon}_k \boldsymbol{R}_k \boldsymbol{\Upsilon}_k^T \right) - 2\boldsymbol{P}_k^- \boldsymbol{H}_k^T - 2\boldsymbol{S}_k^- \boldsymbol{W} \boldsymbol{\gamma}_k^T + 2\boldsymbol{K}_k \boldsymbol{\gamma}_k \boldsymbol{W} \boldsymbol{\gamma}_k^T \quad (33)$$

Setting $\partial J / \partial \boldsymbol{K}_k = 0$ and simplifying the formulation, the analytical gain matrix is given as follows

$$\boldsymbol{K}_k = (\boldsymbol{P}_k^- \boldsymbol{H}_k^T + \boldsymbol{S}_k^- \boldsymbol{W} \boldsymbol{\gamma}_k^T)(\boldsymbol{H}_k \boldsymbol{P}_k^- \boldsymbol{H}_k^T + \boldsymbol{\gamma}_k \boldsymbol{W} \boldsymbol{\gamma}_k^T + \boldsymbol{\Upsilon}_k \boldsymbol{R}_k \boldsymbol{\Upsilon}_k^T)^{-1} \quad (34)$$

**Remark.** The formulation of the gain $\boldsymbol{K}_k$ in Eq. (34) is similar to a well-known form of the conventional Kalman filter and it is a closed-form solution, too.

## 4. Simulation results and analysis

MATLAB/Simulink simulations under the IMU/Radio communication integrated navigation scheme during the Mars entry phase have been carried out to show that the sensitivities of the state estimate errors respected to the uncertain parameter, the perturbations coming from the parameter uncertainty, and the performance of the proposed robust ADEKF.

The simulation conditions and parameters, in which the truth and the initial values are all listed in Table 1 [37]. The initial position of orbiting beacon is (7,855,700 m, −461,800 m, 749,820 m), and the corresponding constant velocity is (66.2 m/s, 2206.4 m/s, −413 m/s). The positions of two surface beacons are (3,300,000 m, 420,000 m, 1,350,000 m) and (3,290,000 m, 570,000 m, 755,000 m), respectively, and their velocities are assumed to zeros.

The gravitational constant of the Mars is $\mu = 42{,}828.29 \times 10^9$ m$^3$/s$^2$. The acceleration bias of IMU is $-0.03$ m/s$^2$. The nominal reference atmospheric density is $\overline{\rho}_0 = 2.0 \times 10^{-4}$ kg/m$^3$ and the nominal reference LDR is $\overline{L/D}_0 = 0.156$. The true percentages of the atmospheric density uncertainty and the LDR uncertainty are assumed to be unknown, which are assumed to subject to uniform distributions in simulations, $c_1 \sim U(-0.15, 0.15)$ and $c_2 \sim U(-0.1, 0.1)$, respectively; and in the model they are assumed that $c_1 = 0$ and $c_2 = 0$. The sensitivity–weighting matrix in the proposed robust ADEKF is given by $W = \text{diag}[0.01, 0.1]$. In the IMU/Radio communication integrated navigation scheme, three possible beacons are selected to provide the two-way range measurements. Initial state error covariance matrix is $\boldsymbol{P}_0 = \text{diag}[10^6, 10^5, 10^{-1}, 10^{-5}, 10^{-5}, 10^{-1}]$, process noise covariance matrix is $\boldsymbol{Q}_k = \text{diag}[10, 1, 10^{-6}, 10^{-10}, 10^{-10}, 10^{-6}]$, and measurement noise covariance matrix is $\boldsymbol{R}_k = \text{diag}[10^{-6}, 10^{-6}, 10^{-6}, 20, 20, 40]$.

The simulation is terminated roughly after 400 s when the supersonic chute is deployed. 5000 simulations have run, the root mean squared error (RMSE) is averaged, and the covariance consistency is tested by the normalized mean error (NME) test [38]. In the simulation, we selected the EKF as the comparing filtering.

Fig. 2 shows the sensitivities of the six navigation state estimate errors of the EKF and the ADEKF respected to the percentage of the atmospheric density uncertainty when $c_1 = 7.5\%$ and $W = \text{diag}[0.01, 0]$; Fig. 3 shows the sensitivities of the six navigation state estimate errors of the EKF and the ADEKF respected to the percentage of the LDR uncertainty when $c_2 = 5\%$ and $W = \text{diag}[0, 0.1]$; and Fig. 4 shows that the above state estimate error values of the ADEKF respected to the two uncertain parameters due to one-sigma errors in their uncertainty when $c_1 = 7.5\%$, $c_2 = -5\%$ and $W = \text{diag}[0.01, 0.1]$. In fact, the sensitivity respected to the uncertainty $c_1$ and $c_2$ represents the sensitivity respected to the atmospheric density and the LDR uncertainty, respectively. In Figs. 2 and 3, subfigures show the sensitivity of the ADEKF respected to the two uncertain parameters. In Fig. 2, it shows that the altitude, velocity, and FPA estimate errors in EKF filtering have great sensitivities respected to the atmospheric density uncertainty than these in ADEKF filtering. In Fig. 3, it can be seen that the longitude and latitude estimate errors have low sensitivities respected to the LDR uncertainty, and the others estimate errors are sensitive to the LDR uncertainty. Obviously, the ADEKF respected to the LDR uncertainty has lower sensitivities than those of the EKF in Figs. 2 and 3. The six state estimate perturbation errors coming from the one-sigma errors of the two uncertain parameters are shown in Fig. 4. From Fig. 4, it is observed that the state estimate perturbation errors coming from the atmospheric density are greater than the errors coming from the LDR after 40 s during Mars entry. It implies that the negative effects of the atmospheric density uncertainty are greater than the LDR uncertainty. The EKF error accumulations of each step lead to gradually diverge, and the ADEKF effectively desensitize the negative effects of the parameter uncertainties.

**Table 1**
Initial conditions and model parameters for integrated navigation.

| Parameters | True | Estimated/Initial |
| --- | --- | --- |
| Initial altitude $r$ | 3518.2 km | 3519.2 km |
| Initial velocity $v$ | 5515 m/s | 5525 m/s |
| Initial FPA $\gamma$ | $-11.8°$ | $-12°$ |
| Initial longitude $\theta$ | $-89.872°$ | $-90.072°$ |
| Initial latitude $\lambda$ | $-28.02°$ | $-28.22°$ |
| Initial azimuth $\psi$ | $5.156°$ | $5.356°$ |
| Ballistic coefficient $B$ | 0.016 m$^2$/kg | 0.0176 m$^2$/kg |

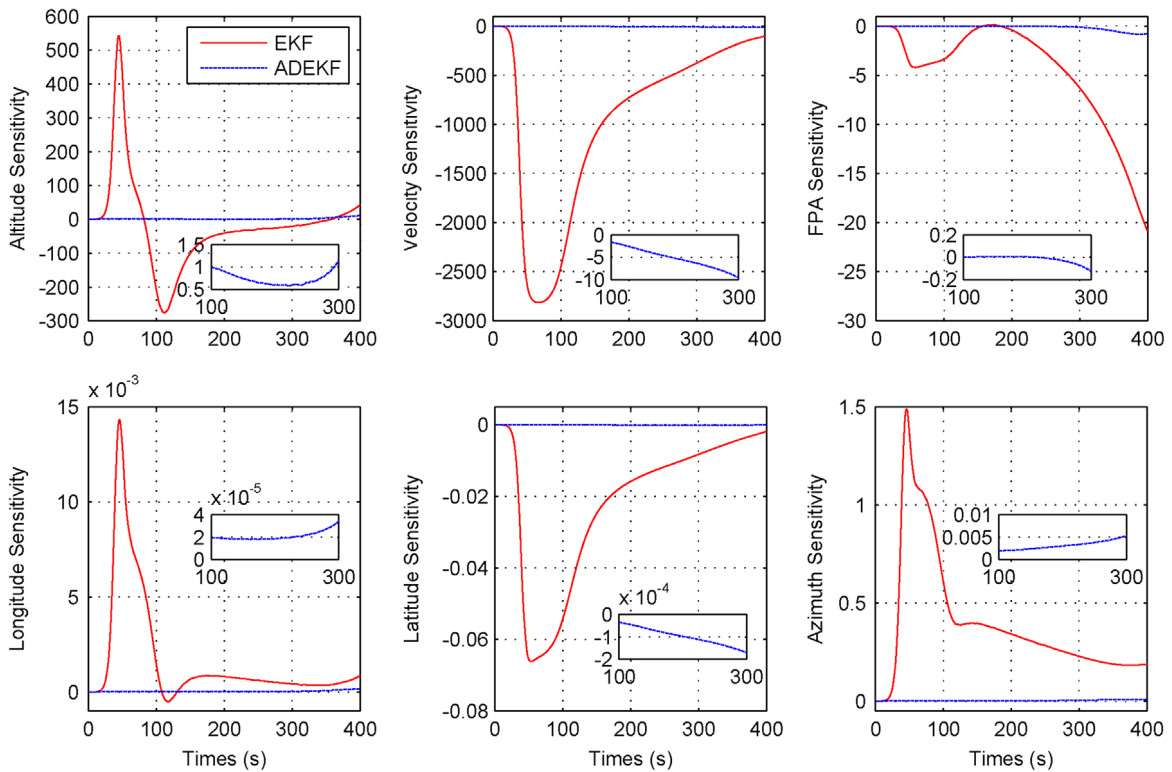

**Fig. 2.** State Sensitivities of EKF and ADEKF respected to the atmospheric density ($c_1 = 7.5\%, c_2 = 0$).

Fig. 5 shows that the results coming from a single run of the Monte Carlo simulations, which are the state estimate errors and $3\sigma$ bounds of ADEKF, when $c_1 = 7.5\%$, $c_2 = -5\%$ and $W = \text{diag}[0.01, 0.1]$. All the six state errors of the ADEKF are captured by the three-sigma covariance bounds for the most parts, and the filter slightly diverges for the FPA and azimuth. Fig. 6 shows that the state estimate RMSEs of the EKF and the ADEKF with logarithmic scales, it can be seen that the RMSEs of the ADEKF are all obviously smaller than these of the EKF except the altitude, and in altitude slightly smaller than these of the EKF.

A filter consistency test is used to highlight the effectiveness of the covariance estimates in accounting for errors in the state estimates [38]. The results of the NME test with logarithmic scales are shown in Fig. 7. The ADEKF covariance estimate accurately predicts the state estimate errors, because all of its state test statistics are well below the required threshold. Obviously, the EKF results imply that it has poor consistency in the altitude, velocity, FPA, longitude and azimuth. It is to say that the ADEKF consistently provide excellent state estimate and the EKF has trouble accounting for the errors with two uncertain parameters.

As a whole, the ADEKF effectively provides a large reduction in sensitivity respected to the two parameter uncertainties, enhances the robustness of integrated navigation, and improve the navigation accuracy.

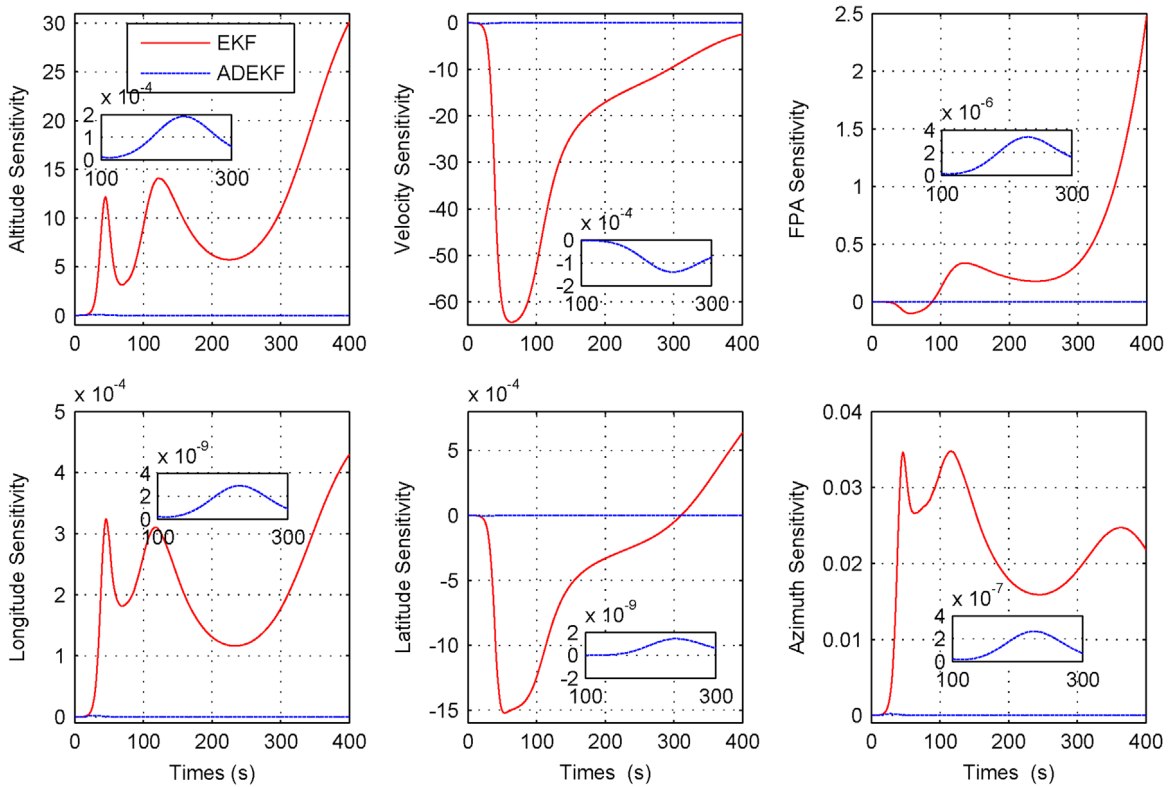

**Fig. 3.** State Sensitivities of EKF and ADEKF respected to the LDR ($c_1 = 0, c_2 = 5\%$).

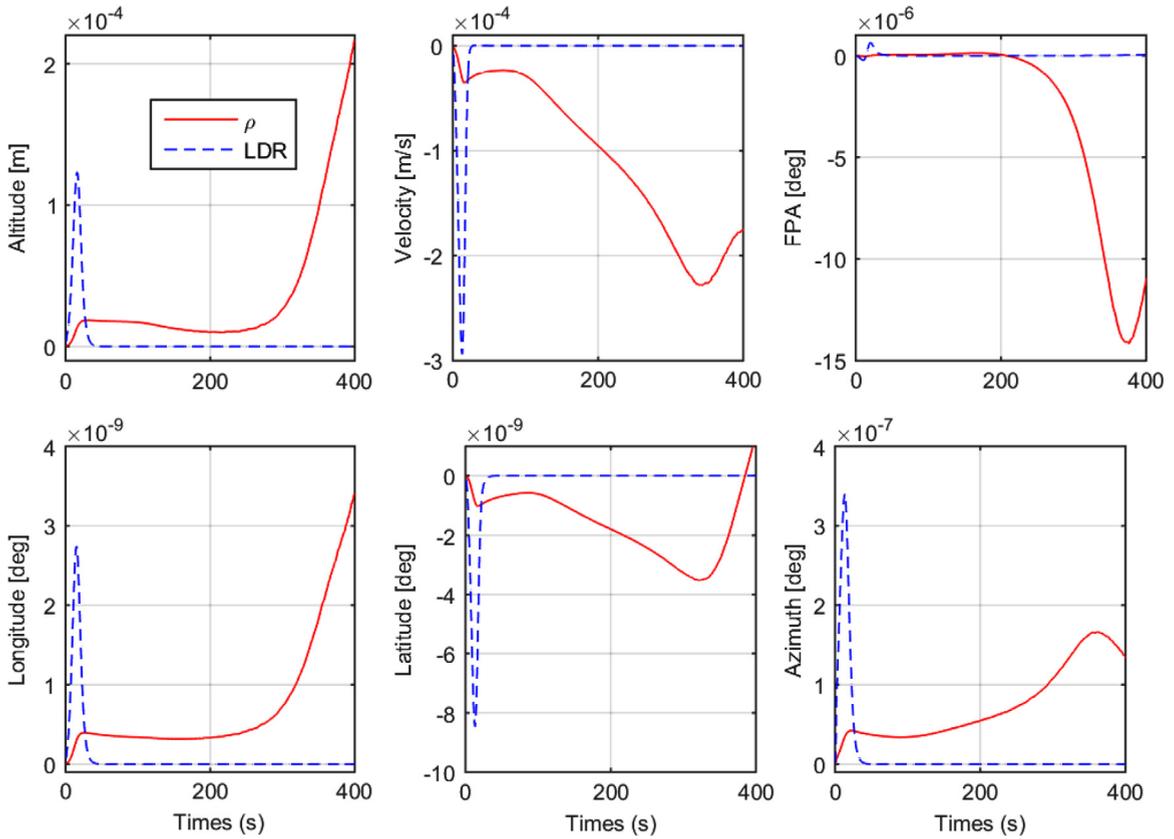

**Fig. 4.** State estimate errors of ADEKF for $1\sigma$ error of the two parameters ($c_1 = 7.5\%, c_2 = -5\%$).

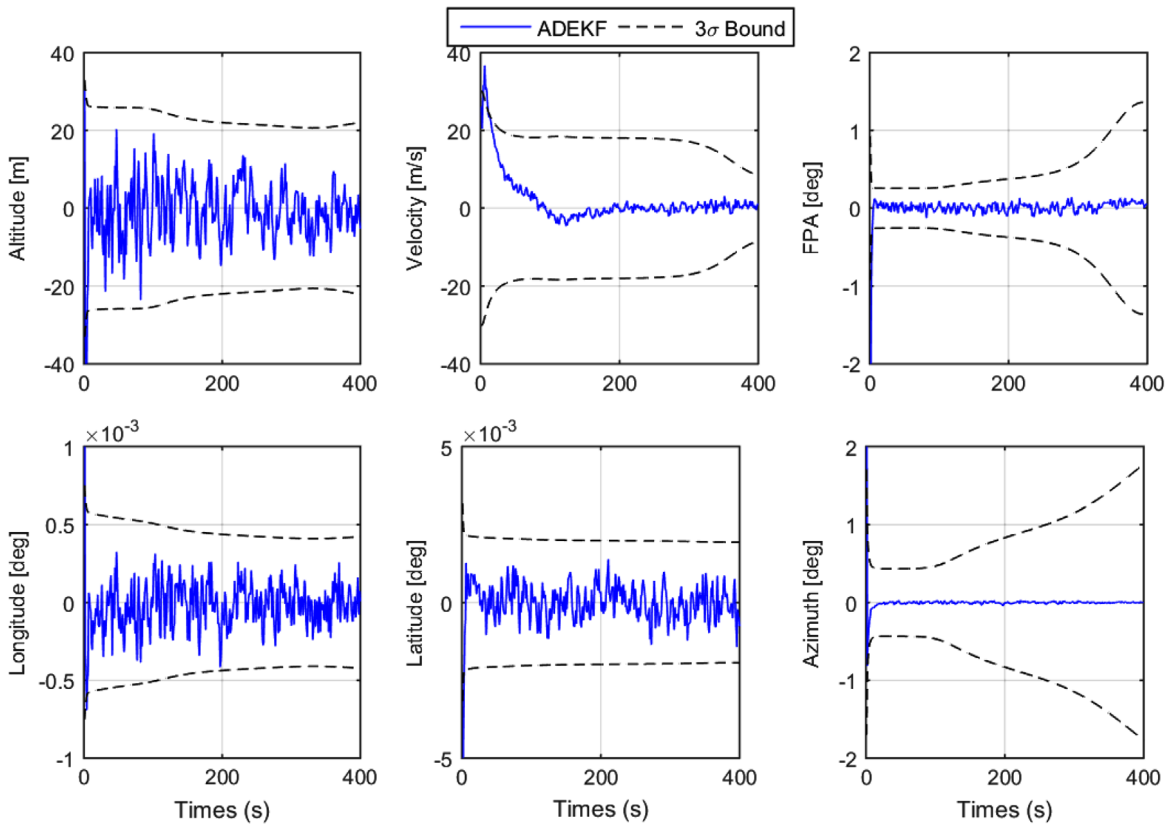

**Fig. 5.** State estimate errors and $3\sigma$ bounds of ADEKF ($c_1 = 7.5\%, c_2 = -5\%$).

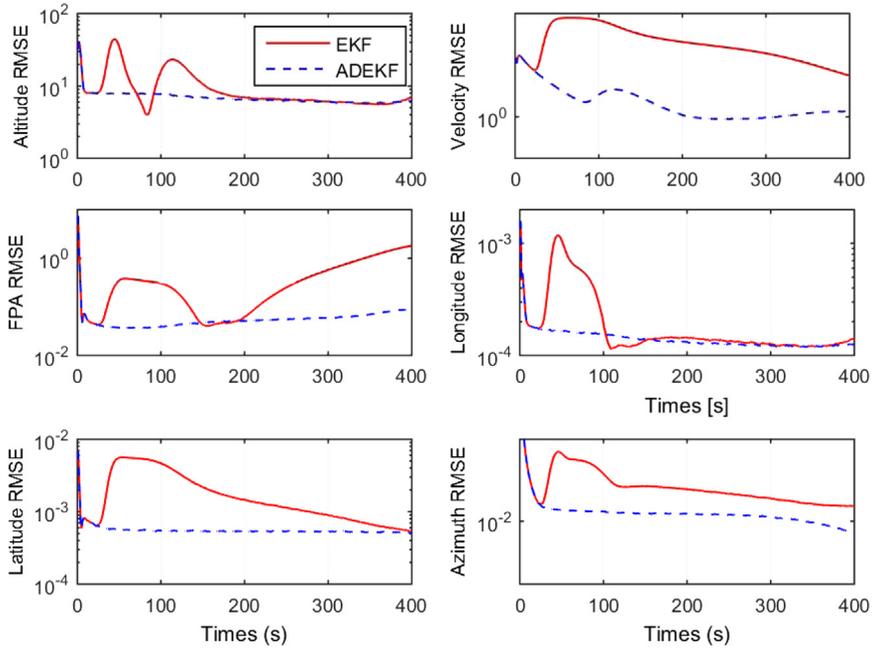

**Fig. 6.** State estimate RMS errors of the EKF and ADEKF.

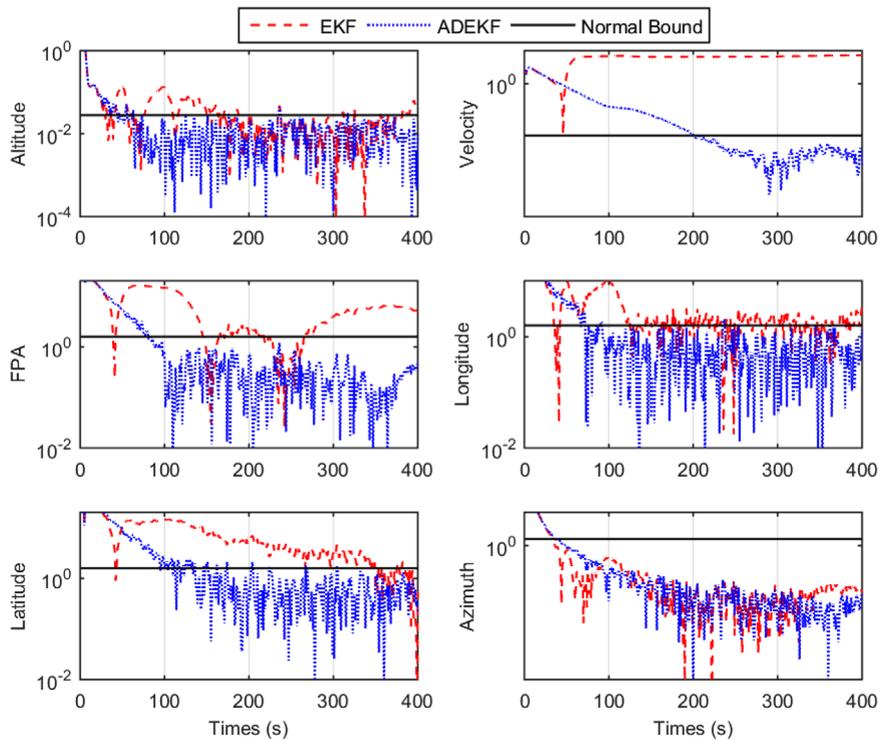

**Fig. 7.** NME test results for state estimate errors.

## 5. Conclusions

High-precision robust atmospheric entry navigation is a key technology to meet the future Mars exploration missions required the pinpoint landing capability. This paper presented a robust integrated navigation algorithm during the Mars atmospheric entry based on a special robust ADEKF. A new cost function penalized by a trace weighted norm of the state error sensitivities is minimized to give a closed-form gain matrix. In the IMU/Radio beacons scheme, the uncertainties of the atmospheric density and the LDR are modeled, and the corresponding sensitivity matrices and perturbation matrices of the vehicle's state estimate errors are employed to describe the effect of their uncertainties. Numerical simulations and the consistency test show that the robust integrated navigation algorithm based on the robust ADEKF effectively reduces the negative effects of the two parameter uncertainties during the Mars entry, and improves the entry navigation robustness and accuracy.

In this work, we give a special sensitivity–weighting matrix in the application during Mars entry. But, how to select and obtain the sensitivity–weighting matrix is an open problem. In the future work, we will focus on how to select a reasonable sensitivity–weighting matrix, which is beyond the scope of the present paper.

### Acknowledgments

The work described in this paper was supported by the National Natural Science Foundation of China (Grant no. 11202023). The authors fully appreciate the financial supports.

### Appendix A. Matrix trace calculus

To get the optimal gain from the cost function in Kalman filter derivations, taking the partial derivative of the trace of matrix is often used. The corresponding results about the derivatives are

$$\frac{\partial}{\partial \boldsymbol{K}} Tr(\boldsymbol{KP}) = \boldsymbol{P}^T \tag{A.1}$$

$$\frac{\partial}{\partial \boldsymbol{K}} Tr\left(\boldsymbol{PK}^T\right) = \boldsymbol{P} \tag{A.2}$$

$$\frac{\partial}{\partial \boldsymbol{K}} Tr\left(\boldsymbol{KPK}^T\right) = \boldsymbol{KP}^T + \boldsymbol{KP} \tag{A.3}$$

where $\boldsymbol{K}$ and $\boldsymbol{P}$ are two arbitrary matrices satisfying matrix multiplication rules.

For the product of a column n-vector $\boldsymbol{x}$ with itself, the traces of them satisfy

$$Tr(\boldsymbol{xx}^T) = \boldsymbol{x}^T \boldsymbol{x} \tag{A.4}$$